\documentclass[newstyle,twocolumn,journal]{rmaa}
\usepackage{paralist}

\usepackage{psfrag,color}

\usepackage[latin1]{inputenc}

\title{The Parker instability in axisymmetric filaments: 
final equilibria with longitudinal magnetic field}
\author{F.~J.~S\'anchez-Salcedo\altaffilmark{1}
          and A.~Santill\'an\altaffilmark{2}}
\altaffiltext{1}{Instituto de Astronom\'\i a, Universidad Nacional Aut\'onoma
de M\'exico, Ciudad Universitaria, 04510, Mexico City, Mexico.}
\altaffiltext{2}{C\'omputo Aplicado-DGSCA, Universidad Nacional Aut\'onoma
de M\'exico, Ciudad Universitaria, 04510, Mexico City, Mexico.}

\shortauthor{S\'anchez-Salcedo \& Santill\'an}
\shorttitle{The Parker instability in filaments}
\fulladdresses{
\item F. Javier S\'anchez-Salcedo: Instituto de Astronom\'{\i}a,
Universidad Nacional Aut\'onoma de M\'exico,
  Apartado Postal 70-264, Ciudad Universitaria, Mexico City, Mexico
  (jsanchez@astroscu.unam.mx).
\item Alfredo Santill\'an: C\'omputo Aplicado,
Direcci\'on General de Servicios de C\'mputo Acad\'emico,
Universidad Nacional Aut\'onoma de M\'exico, Ciudad Universitaria,
Mexico City, Mexico (alfredo@astro.unam.mx).}
\listofauthors{F.~J. S\'anchez-Salcedo \& A. Santill\'an}
\indexauthor{S\'anchez-Salcedo, F.~J.} 
\indexauthor{Santill\'an, A.}
\resumen{Se estudian los estados finales de
equilibrio que surgen de la inestabilidad de Parker cuando partimos 
de una configuraci\'on cil\'{\i}ndrica de gas en equilibrio
magnetohidrost\'atico en un campo gravitacional radial y 
con un campo magn\'etico longitudinal.
Nuestro objetivo es comparar
los estados de equilibrio no lineales con los que se obtienen 
en los sistemas con geometr\'{\i}a Cartesiana. 
Se presentan los mapas de densidad y de las l\'{\i}neas de campo
magn\'etico en ambas geometr\'{\i}as para un campo gravitacional
de intensidad constante. 
Encontramos que el flotamiento magn\'etico es menos
eficiente bajo simetr\'{\i}a axial que en una atm\'osfera
Cartesiana. Como consecuencia, las condensaciones que se forman
en el modelo axisim\'etrico tienen menor densidad columnar.
Por ende, el cociente entre la presi\'on magn\'etica y t\'ermica 
en el estado final toma valores m\'as extremos bajo simetr\'{\i}a
Cartesiana. Se discuten tambi\'en algunos modelos en los que el
campo gravitacional no es uniforme.
    }

\abstract{We study the final equilibrium states of the Parker instability
arising from an initially unstable cylindrical equilibrium
configuration of gas in the presence of a radial gravitational field
and a longitudinal magnetic field. 
The aim of this work is to compare the properties of the
nonlinear final equilibria with those 
found in a system with Cartesian geometry. 
Maps of the density and magnetic field lines, when the strength of the
gravitational field is constant, are given in both geometries.
We find that the magnetic buoyancy and the drainage of
gas along field lines are less efficient under axial symmetry
than in a Cartesian atmosphere. 
As a consequence, the column density enhancement arising in gas
condensations in the axially-symmetric model is smaller than in 
Cartesian geometry.
The magnetic-to-gas pressure ratio in the final state takes
more extreme values in the Cartesian model. Models with non-uniform
radial gravity are also discussed.
}


\begin{document}
\maketitle

\section{Introduction}
Parker (1966) demonstrated that a layer
that is supported against the vertical gravity by
the pressure of interstellar gas, horizontal magnetic field and cosmic rays,
is always unstable to long wavelength deformations of the field lines 
because of the buoyancy of magnetic
field and cosmic rays. Much work has been done to understand the role
of the Parker instability in modeling the structure of
the interstellar medium (ISM) and, more specifically, its role in gathering gas
to trigger the formation of giant molecular clouds (e.g.,
Mouschovias et al.~1974; Blitz \& Shu 1980; Mouschovias
et al.~2009).  Giz \& Shu (1993), 
Kim et al.~(1997) and Kim \& Hong (1998) examined how the non-uniform 
nature of the Galactic gravity
might affect the length and timescales of the Parker instability.
Many studies have been devoted to investigate the effect of other
physical ingredients on the Parker instability 
(Shu 1974; Nakamura et al.~1991; Hanawa et al.~1992; Hanasz \& Lesch 2000;
Kim et al.~2000; Santill\'an et al.~2000; Kim et al.~2001; Franco et al.~2002;
Kosi\'{n}ski \& Hanasz 2006, 2007; Lee \& Hong 2007; Mouschovias et al.~2009).

\begin{figure}
  \includegraphics[width=80mm,height=85mm]{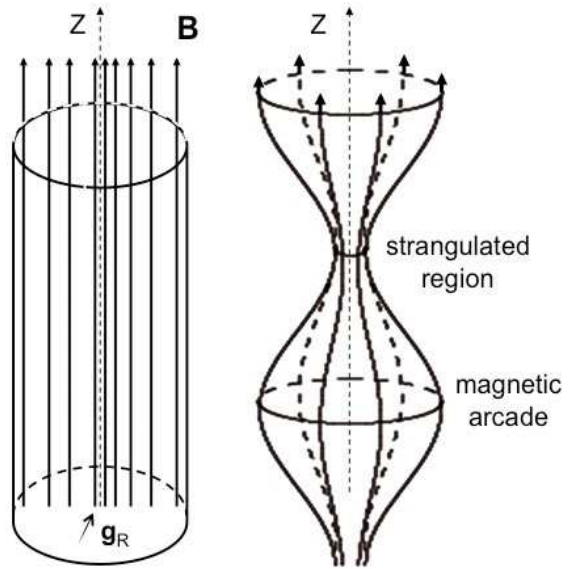}
\vskip 0.4cm
  \caption{
Sketch of the magnetic field configuration in the initial (left
panel) and final (right panel) equilibrium
states in axial symmetry.
  }
  \label{fig:sketch}
\end{figure}

The structure of the final static state resulting from
the Parker instability was derived by Mouschovias (1974, hereafter M74)
in the Cartesian model, namely,
for a non-rotating layer under a vertical gravity, with a horizontal magnetic
field in the initial equilibrium state. 
In the unperturbed state, the gas is in hydrostatic equilibrium 
in a stratified layer,
so that all the quantities depend only on the distance from
the midplane $y$, and the magnetic field is plane-parallel 
and runs along the $x$-direction. In this {\it two-dimensional Cartesian}
problem, one assumes that all the quantities are independent of the third
dimension, $z$, at any time.
In the final state, the magnetic field rises in certain regions and
sink in others forming bulges and valleys.
Matter loads onto the field lines by draining down from the top region of the
bulge and sinking into the valleys. 
So far, the final states of the Parker instability have been discussed 
in Cartesian models. In various astrophysical systems, however, 
the gas tends to form structures with other symmetries, such as filaments
or elongated structures (e.g., Genzel \& Stutzki 1989; Alfaro et al.~1992;
Ryu et al.~1998; Conselice et al.~2001; Salom\'e et al.~2006; 
Jackson et al.~2010). 
Suppose now that we have a very elongated
self-gravitating filamentary structure with a magnetic field 
traveling along the major
axis of this structure, such that it has the effective geometry
of a infinite cylinder. If the filament is supported against the radial
self-gravity of the system by the pressure of gas and the longitudinal
magnetic field, long wavelengths 
deformations of the magnetic field lines will allow matter to slide radially
along field lines under the action of the gravitational field towards
the symmetry axis, whereas the magnetic field tends to expand in
radius, through a process of magnetic buckling (see Fig.~\ref{fig:sketch}). 
If the disturbances are axisymmetric, we expect the alternate
formation of magnetic bottlenecks in the condensations of gas separated 
by portions of inflated magnetic lines or arcades, resembling a sausage mode. 
The final state is expected to be different than in a vertically
stratified disk as those models currently used to study the Parker
instability in galactic disks. In fact, the response of the system
and the final equilibrium state depend on the adopted model:
Cartesian or axisymmetric.

In this paper, we study the nonlinear final equilibria of the
Parker instability of a non-rotating, cylindrical initial state.
In the initial state, the magnetic field is assumed to be parallel
to the axis of the cylinder, which is taken to coincide with the
$z$-axis. The cylinder is supported against the action of a radial
gravitation field $g_{R}$ by the thermal and magnetic pressures 
(see Fig.~\ref{fig:sketch}).
We will assume that
the gas is isothermal and evolves under flux-freezing conditions. 
For axisymmetric perturbations, all the quantities only depend
on $z$ and $R$ (not on the azimuthal angle $\phi$) and hence 
the problem is also two-dimensional.
The evolution of the Parker instability
in the $(z,R)$-plane is expected to be qualitative analogous in the
main features, to the evolution found in the $(x,y)$-plane for the case of a 
plane-parallel initial state by M74.
Throughout the paper, we will use the coordinate $z$ to denote
the distance on the axis of symmetry in the axisymmetric model and
should not be confused with the third component of the Cartesian model.

The aim of this work is to compare in detail
how the final distributions of density and magnetic field vary
from Cartesian to axisymmetric geometry\footnote{Note that the change 
in the geometry of the system is, by no means, equivalent to
a coordinate transformation.}.
Accurate predictions of the nonlinear final states, an interesting
problem in itself, can be very useful for the purpose of testing MHD
codes.
We will follow Mouschovias' procedure (1974, 1976) to construct
the final equilibrium states.
Nakano (1979) and Tomisaka et al.~(1988) used the
same technique to find the structure of axisymmetric magnetized clouds 
that can be reached from {\it nonequilibrium spherical} configurations.
By contrast, we start from a cylindrical configuration in magnetohydrostatic
equilibrium as those used to study the gravitational collapse and
fragmentation of filamentary clouds (e.g., Nakamura et al.~1995). 
In these simulations, however, the gravitational instabilities
play a dominant role and not the Parker instability itself.
In order to isolate the outcomes of the Parker instability,
we will not consider self-gravity of the gas.
In addition, we will also ignore the destabilizing influence of
the interchange mode and the magnetic field curvature  
in the equilibrium states (Ass\'eo et al.~1978, 1980; 
Lachi\`eze-Rey et al.~1980), which could lead to the development
of substructure. The study of these instabilities requires a
full magnetohydrodynamical model and they will not be discussed 
further here.

Our paper is organized as follows. In Section \ref{sec:initialstates}, 
we describe
the initial equilibrium configurations. The elliptic differential
equations that give the equilibrium states under flux-freezing conditions
are provided in Section \ref{sec:equilibrium}. In 
Section \ref{sec:final} we compare
the resulting features of the final states arising from the Parker
instability in Cartesian and axisymmetric models. A summary of the results
are presented in Section \ref{sec:conclusions}.

\section{Initial states}
\label{sec:initialstates}
We consider first the initial state in the Cartesian model, 
that is, a planar layer
that is supported against the vertical gravity field by the gas pressure $P$
and a horizontal magnetic field. We will use a Cartesian coordinate system
$(x,y)$, where $x$ is the horizontal coordinate and 
$y$ is the distance to the midplane of the layer,  
so that $\vec{B}_{0}=(B_{0,x}(y),0)$ in
the initial state.
Throughout the paper, the subscript $0$ signifies the initial state.
Magnetohydrostatic equilibrium is satisfied if
\begin{equation}
\vec{\nabla}\left(P_{0}+\frac{B_{0,x}^2}{8\pi}\right)=-\rho \vec{\nabla}\psi,
\end{equation}
where $\psi=\psi(y)$ is the gravitational potential. 
If the gas is isothermal with thermal
sound speed $c_{s}$ and the ratio of the magnetic
to gas pressures, $\alpha\equiv B_{0}^{2}/8\pi P_{0}$, 
is constant, the equilibrium equation reads
\begin{equation}
(1+\alpha)c_{s}^{2}\vec{\nabla} \ln \rho_{0} =- \vec{\nabla}\psi,
\end{equation}
which has the solution
\begin{equation}
\rho_{0}(y)=\rho_{0}(0) \exp \left(-\frac{\psi}{(1+\alpha)c_{s}^{2}}\right),
\end{equation}
with $\psi(0)=0$.
If the gravitational field $\vec{g}=-\vec{\nabla}\psi$ 
is taken to be vertical and constant 
(but it reverses its direction across the midplane), 
we have $\psi=g|y|$, with $g={\rm constant}>0$.
In this particular case, the mass density decays exponentially with
a typical scale $H$ of $(1+\alpha) c_{s}^{2}/g$, whereas the magnetic
field strength decays with a characteristic scale $2H$.

If, instead of a planar atmosphere, we assume that the
gas extends to infinity along a cylinder whose axis coincides
with the $z$-axis, and is threaded by an axial magnetic field $B_{z}$,
the density profile at equilibrium is: 
\begin{equation}
\rho_{0}(R)=\rho_{0}(0) \exp \left(-\frac{\psi}{(1+\alpha)c_{s}^{2}}\right),
\end{equation}
where $R$ is the radial distance in cylindrical coordinates 
and $\psi(R)$ is the
gravitational potential, with $\psi(0)=0$. In fact,
we assume that the gravitational acceleration has only a radial
component. 
If the radial gravitational force is taken to be constant, it holds
that $\psi=gR$. Therefore, the radial scalelength for the density is
$H=(1+\alpha) c_{s}^{2}/g$, as in Cartesian geometry.

We see that if the gravitational potential $\psi(y)$
in the infinite planar layer is identical to  
the gravitational potential $\psi(R)$ in the cylindrical model,
the profiles of density, magnetic field and gas pressure along $y$ and $R$ 
will be identical in both models.
However, these systems are no longer equivalent when the Parker instability
arises. In the Cartesian model, magnetic field lines are allowed to
inflate only in the vertical direction, whereas the expansion of field
lines as well as the contraction of matter occur radially in the axisymmetric
model (see Fig.~\ref{fig:sketch}). 
The question that arises is how the distributions of density,
pressure and field lines in an azimuthal section [i.e.~in the $(z,R)$-plane] 
look like as compared to
those found in a vertical section [i.e.~in the $(x,y)$-plane] 
in the well-studied Cartesian model.

When studying the Parker instability at the solar neighborhood, it is
common to use a local Cartesian frame where the azimuthal direction
is identified with the horizontal coordinate $x$ and the vertical
direction of the Galaxy with $y$. In the two-dimensional approximation,
the third radial direction is ignored. As a cautionary comment, 
our coordinate $R$ used throughout the paper should not
be confused with the radial direction of the Galaxy.

\section{The equilibrium equations}
\label{sec:equilibrium}
If a disturbance of sufficiently long wavelength 
is applied along the initial magnetic field
to the initial equilibrium states described in the previous Section,
the Parker instability develops in time and parts of the magnetic field
lines bulge outward until the tension of the field lines increases
sufficiently for an equilibrium state to be established. In the final
states arising from the Parker instability, the magnetic field lines are
curved and the magnetohydrostatic force equation is
\begin{equation}
-\vec{\nabla}P-\rho\vec{\nabla}\psi +\frac{1}{c}\vec{J}\times \vec{B}=0,
\label{eq:primitive}
\end{equation}
where $\vec{J}$ is the current density.
The coordinate symmetries of the unperturbed states 
(see Section \ref{sec:initialstates})
reduce the problem to the $(x,y)$-plane in the Cartesian model
and to the $(z,R)$-plane in the axisymmetric model.
Under these symmetries, flux-freezing allows to transform the 
magnetostatic equations of equilibria into a second-order elliptic
partial differential equation (Dungey 1953), conventionally called
the Grad-Shafranov equilibrium equation.
The beauty of this approach is that the final equilibrium
states can be found with no need to solve a time-dependent 
problem (M74; Mouschovias 1976). 
In the following we give the Grad-Shafranov equilibrium equations
and the boundary conditions.
Details on its derivation can be found in M74, Mouschovias (1976) and
Nakano (1979).

\subsection{The Cartesian model}
\label{sec:Cartesian}
In the Cartesian layer with one ignorable coordinate, so that all
quantities depend only on $(x,y)$,
the magnetic vector potential can be written as 
${\vec{A}}=A(x,y)\hat{e}_{z}$ and then
$B_{x}=\partial A/\partial y$ and $B_{y}=-\partial A/\partial x$.  
One can show that the scalar function $q(x,y)$, 
defined by
\begin{equation}
q=P\exp\left(\frac{\psi}{c_{s}^{2}}\right),
\end{equation}
is constant on a field line {\it at magnetohydrostatic equilibrium}, 
and is given by:
\begin{equation}
q(A)=\frac{c_{s}^{2}}{2}\frac{d\tilde{m}}{dA} \left[\int_{0}^{\lambda_{x}/2}
\!\!\!\!dx \frac{\partial y(x,A)}{\partial A} \exp\left(-\frac{\psi(x,A)}{c_{s}^{2}}
\right)\right]^{-1},
\label{eq:qA}
\end{equation}
where $\lambda_{x}$ is the perturbation wavelength along the initial
magnetic field and
$d\tilde{m}$ is the mass per unit length (along the ignorable coordinate) 
in a flux tube between field lines characterized by
$A$ and $A+dA$ and between $x=0$ and $x=\lambda_{x}$ 
(i.e.~the mass-to-flux ratio). 
Under flux-freezing conditions, the mass-to-flux ratio 
is a constant of motion and, therefore, can be determined from the initial
equilibrium configuration:
\begin{equation}
\frac{d\tilde{m}}{dA}=\lambda_{x}\frac{\rho_{0}(A)}{B_{0}(A)}.
\end{equation}

Finally, the magnetic equilibrium equation (\ref{eq:primitive}) may be written 
in terms of $A$ as
\begin{equation}
\nabla^{2} A=-4\pi \frac{dq}{dA} \exp\left(-\frac{\psi}{c_{s}^{2}}\right)
\label{eq:ellipticxy}
\end{equation}
(Dungey 1953; M74).
Once the boundary conditions are specified,
Equations (\ref{eq:qA}) and (\ref{eq:ellipticxy}) can be solved 
numerically by using a iterative scheme (M74).

The boundary conditions we take are the same as those in M74.
The system is assumed to be periodic in $x$ and symmetric about the $x$-axis
so that Equation (\ref{eq:ellipticxy}) is solved in the rectangle 
$0<x<X_{\rm max}$ with $X_{\rm max}=\lambda_{x}/2$ and
$0<y<Y_{\rm max}$, with the boundary conditions
\begin{equation}
\frac{\partial A}{\partial x}\bigg|_{x=0, X_{\rm max}}=0.
\end{equation}
The upper boundary must be far enough from the midplane
not to affect the evolution since a small vertical size of the computational
domain suppresses the instability (e.g., M74; Mouschovias 1996). 
At the upper boundary we impose that the field lines are
not deformed, $A(x,Y_{\rm max})=A_{0}(Y_{\rm max})$.
Due to the imposed reflection symmetry about the midplane,
the field line originally coinciding with the $x$-axis remains
undeformed.

In the following, we provide the equations for the particular
case of a constant external field.
Although a uniform acceleration is
only realistic at high altitudes from the midplane,
we will concentrate on this
case for simplicity and to facilitate comparison with some previous
work that use a constant gravitational field (e.g., M74; Mouschovias
et al.~2009). A realistic model of the gravitational potential
is important to derive the growth rate of the instability and
the most unstable wavelength, which determines the separation of the
magnetic valleys (Kim \& Hong 1998; Section \ref{sec:nonuniform}). 
For now, we are mainly interested in
the main conceptual features that arise in the final equilibria state
when the geometry of the problem is changed, even if the adopted
spatial dependence of the 
gravitational field lacks a strong astrophysical motivation.

As said in Section \ref{sec:initialstates}, 
if the external gravitational field is constant,
the gravitational potential is $\psi=g|y|$, with $g$ a positive
and constant acceleration. 
The equations are put in dimensionless form by choosing the following
natural units: $c_{s}^{2}/g$, $c_{s}$ and $\rho_{0}(0)$, 
for length, velocity and density, respectively,
The magnetic vector potential will be measured
in units of $-2HB_{0}(0)$. 
Note that the dimensionless vertical scale is $H=(1+\alpha)$ in these
units.  The dimensionless of Equation (\ref{eq:ellipticxy})
has the form
\begin{equation}
\nabla^{2} A= \hat{Q}_{xy}(x,y;\alpha),
\label{eq:ellipticxyadim}
\end{equation}
where
\begin{equation}
\hat{Q}_{xy}=-\frac{1}{8\alpha (1+\alpha)^2}\frac{dq}{dA}\exp (-y).
\label{eq:Qxy}
\end{equation}
In these units,
\begin{equation}
q(A)= \frac{1}{2} \frac{d\tilde{m}}{dA}\left[\int_{0}^{\lambda_{x}/2}
dx \frac{\partial y(x,A)}{\partial A} \exp[-y(x,A)]\right]^{-1},
\end{equation}
where 
\begin{equation}
\frac{d\tilde{m}}{dA}=-2 \lambda_{x} (1+\alpha) A.
\end{equation}

Under constant gravity condition, 
the vector potential in the plane-parallel unperturbed state 
is
\begin{equation}
A_{0}(y)= \exp\left(-\frac{y}{2(1+\alpha)}\right),
\label{eq:A0}
\end{equation}
and the source term of Eq.~(\ref{eq:ellipticxyadim}) is
\begin{equation}
\hat{Q}_{xy}^{(i)}= \frac{1}{4(1+\alpha)^{2}} \exp\left(-\frac{y}{2(1+\alpha)}\right).
\label{eq:rhs_2D}
\end{equation}
In this initial state, the relation between $A$ and $q$ is
$A_{0}(q)=q^{-1/(2\alpha)}$ (see also M74).

\subsection{The axisymmetric model}
\label{sec:axisymmetric}
Now, we consider a three-dimensional geometry with axial symmetry
about the $z$-axis and longitudinal magnetic field ($B_{\phi}=0$). 
The magnetic field is related to the potential
vector $\vec{A}=A(z,R)\hat{e}_{\phi}$ 
by $B_{z}=R^{-1} \partial (RA)/\partial R$ and 
$B_{R}=-\partial A/\partial z$.
Under axisymmetric disturbances, the magnetic flux $\Phi(R,z)\equiv R A$ is
constant on a magnetic surface and is also a constant of the motion.
In this geometry, the function $q$ defined in Eq.~(\ref{eq:qA}) 
is given by
\begin{equation}
q(\Phi)=\frac{c_{s}^{2}}{2\pi}\frac{dm}{d\Phi} \left[\int_{0}^{\lambda_{z}/2}
\!\!\!\!dz \frac{\partial R^{2}(z,\Phi)}{\partial \Phi} \exp\left(-\frac{\psi(z,\Phi)}{c_{s}^{2}}
\right)\right]^{-1},
\label{eq:qPhi}
\end{equation}
where $\lambda_{z}$ is the longitudinal perturbation wavelength
and $dm$ is the mass in a flux tube between $\Phi$ and $\Phi+d\Phi$
and between $z=0$ to $z=\lambda_{z}$.
Again, $q$ is a constant along field lines in equilibrium configurations.
Under flux-freezing conditions,
the mass-to-flux ratio is a constant of the motion, which
can be determined from the initial configuration by:
\begin{equation}
\frac{dm}{d\Phi}=2\pi \lambda_{z}\frac{\rho_{0}}{B_{0}}.
\end{equation}
The balance of forces (Eq.~\ref{eq:primitive}) in the final magnetic
configuration can be rewritten in terms of $\Phi$ as
\begin{equation}
\nabla^{2}\Phi - \frac{2}{R}\frac{\partial \Phi}{\partial R}= -4\pi R^{2}\frac{dq}{d\Phi}
\exp\left(-\frac{\psi}{c_{s}^{2}}\right)
\label{eq:ellipticzR}
\end{equation}
(Mouschovias 1976; Nakano 1979).
Equation (\ref{eq:ellipticzR}) together with Eq.~(\ref{eq:qPhi}) 
enable us to derive $\Phi(z,R)$ once the boundary conditions are specified.
The magnetic field can be found just by simple derivation: 
$B_{z}=R^{-1}\partial \Phi/ \partial R$ and $B_{R}=-R^{-1}\partial \Phi/ \partial
z$.

The computational domain is a rectangular box of $0<z<Z_{\rm max}$ and
$0<R<R_{\rm max}$. 
Because of the symmetries of the initial and final states,
the length of the computation box in the $z$-direction, $Z_{\rm max}$,
is taken to be $\lambda_{z}/2$, half the wavelength of the perturbation.
Periodicity in $z$ is translated to 
\begin{equation}
\frac{\partial \Phi}{\partial z}\bigg|_{z=0, Z_{\rm max}}=0.
\end{equation}
The definition of $\Phi$ means that $\Phi(0,z)=0$ at any time.
Assuming that the deformation of field lines can be neglected  
on the surface of the cylinder, we impose that
$\Phi(z,R_{\rm max})=\Phi_{0} (R_{\rm max})$. 

\begin{figure}
  \includegraphics[width=1.6\columnwidth]{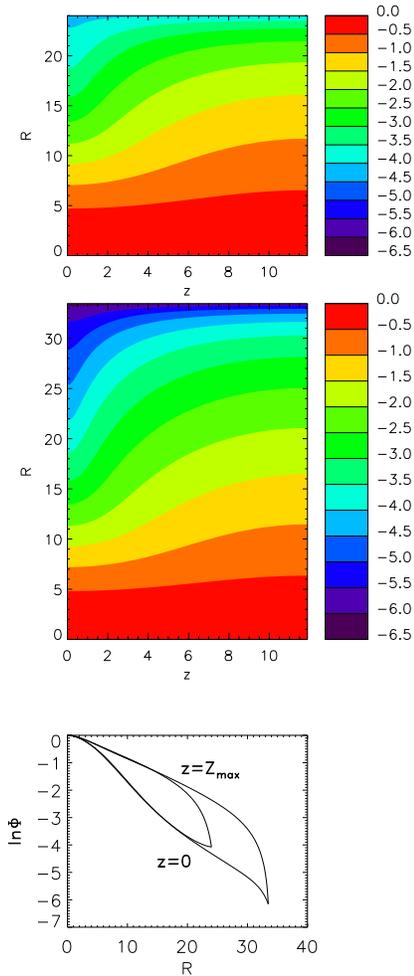}
\vskip 0.4cm
  \caption{
Azimuthal section of the magnetic flux $\Phi(z,R)$ 
in the axially-symmetric final state for $\alpha=1$ and $g_{R}=g={\rm const}$,
when the upper boundary is placed at
$R_{\rm max}=24$ (top) and at $R_{\rm max}=33.5$ (middle). 
Magnetic field lines coincide with the isocontours of $\Phi$.
The color bar is in
natural logarithmic scale. The unit length is $c_{s}^{2}/g$.
A comparison of the radial profile of the magnetic flux along cuts 
at $z=0$ and $z=Z_{\rm max}$ is given in the lower panel when
the boundary is at $R_{\rm max}=24$ and $R_{\rm max}=33.5$.

  }
  \label{fig:hbox}
\end{figure}

\begin{figure}
  \includegraphics[width=80mm,height=85mm]{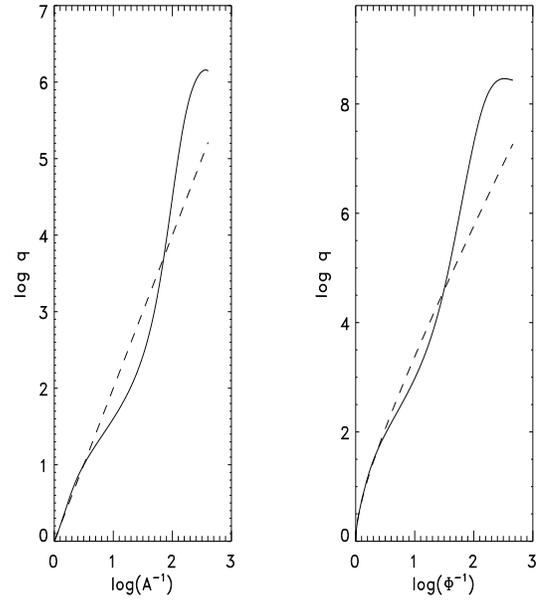}
\vskip 0.4cm
  \caption{
The shape of the function $q(A)$ (left panel) and $q(\Phi)$
(right panel) in the initial state (dashed lines) and in the
final state (solid lines), with $\alpha=1$ and constant gravity, 
for $X_{\rm max}=15$ and $Z_{\rm max}=15$,
respectively. Note that the range of the $q$-axis is different.
  }
  \label{fig:qA}
\end{figure}

The equivalent condition of constant gravity is to assume
that the radial gravitational field is constant,
i.e.~$\vec{g}=-g\hat{e}_{R}$, with $g$ a positive constant 
acceleration.
We may use the same units for distance, velocity and density as those
described in the previous subsection. The  
magnetic flux will be measured in units of $-4H^{2}B_{0}(0)$.
Equation (\ref{eq:ellipticzR}) in dimensionless form is:
\begin{equation}
\nabla^{2}\Phi - \frac{2}{R}\frac{\partial \Phi}{\partial R}= 
Q_{zR}(z,R;\alpha),
\label{eq:lessPhi}
\end{equation}
with
\begin{equation}
Q_{zR}= -\frac{R^{2}}{32\alpha (1+\alpha)^{4}}\frac{dq}{d\Phi}
\exp\left(-R\right),
\label{eq:Qcyl}
\end{equation}
and
\begin{equation}
q(\Phi)=\frac{1}{2\pi}\frac{dm}{d\Phi} \left[\int_{0}^{\lambda_{z}/2}
\!\!\!\!dz \frac{\partial R^{2}(z,\Phi)}{\partial \Phi} \exp\left[-R(z,\Phi)
\right]\right]^{-1},
\label{eq:qPhiadim}
\end{equation}
where the dimensionless mass-to-flux is 
\begin{eqnarray}
&&\frac{dm}{d\Phi}=-8\pi (1+\alpha)^{2} \lambda_{z} \frac{\rho_{0}}{B_{0}}\\
&&= -8\pi (1+\alpha)^{2} \lambda_{z} \exp \left(-\frac{R(\Phi)}{2(1+\alpha)}\right).
\end{eqnarray}
By comparing Eqs.~(\ref{eq:lessPhi})-(\ref{eq:Qcyl}) with
Eqs.~(\ref{eq:ellipticxyadim})-(\ref{eq:Qxy}), we see that the dynamical
variable $\Phi$ also satisfies a nonlinear elliptical differential equation, 
but with a different differential operator and source term than
the equation governing $A$ in the Cartesian model.

In order to gain some physical insight, it is useful to compute the function
$q(\Phi)$ in the initial equilibrium configurations which,
of course, are solutions of the differential equation (\ref{eq:lessPhi}).
In a model with uniform radial gravitational field,
the magnetic flux and the $q$-function in the unperturbed state are, 
respectively:
\begin{equation}
\Phi_{0}(R)=\left(1+\frac{R}{2(1+\alpha)}\right)\exp\left(-\frac{R}{2(1+\alpha)}\right),
\end{equation}
and
\begin{equation}
q_{0}(R)=\exp\left(\frac{\alpha R}{1+\alpha}\right).
\end{equation}
At small enough $R$, say $R\ll 2(1+\alpha)$, they can be
approximated by $\Phi_{0}(R)\approx
\exp (-R/2[1+\alpha])$ and $q_{0}(\Phi_{0})=\Phi_{0}^{-2\alpha}$. 
Interestingly, they have exactly the same forms as $A_{0}(y)$ 
and $q_{0}(A)$ in the Cartesian model, respectively
(see Eq.~\ref{eq:A0} and the very end of Section \ref{sec:Cartesian}).
At large radii, $R\gg 2(1+\alpha)$, or equivalently $\ln q\gg 2\alpha$, 
the relationship between $\Phi_{0}$
and $q_{0}$ is $\Phi_{0} \simeq (2\alpha)^{-1}  q_{0}^{-1/(2\alpha)}\ln q_{0}$.
Therefore, at these large $q$ values, $q$ increases 
with $\Phi^{-1}$ slightly faster than the rate at which $q$ 
increases with $A^{-1}$ in the Cartesian model 
($q=A^{-2\alpha}$). In Section \ref{sec:final}, we will calculate the
dependence of the function $q$ on $\Phi$ in the final axially-symmetric
equilibrium state. The shape of $q(\Phi)$ as compared to
$q(A)$ in the Cartesian model will be also discussed.

\begin{figure*}
  \includegraphics[viewport=-30 0 264 279]{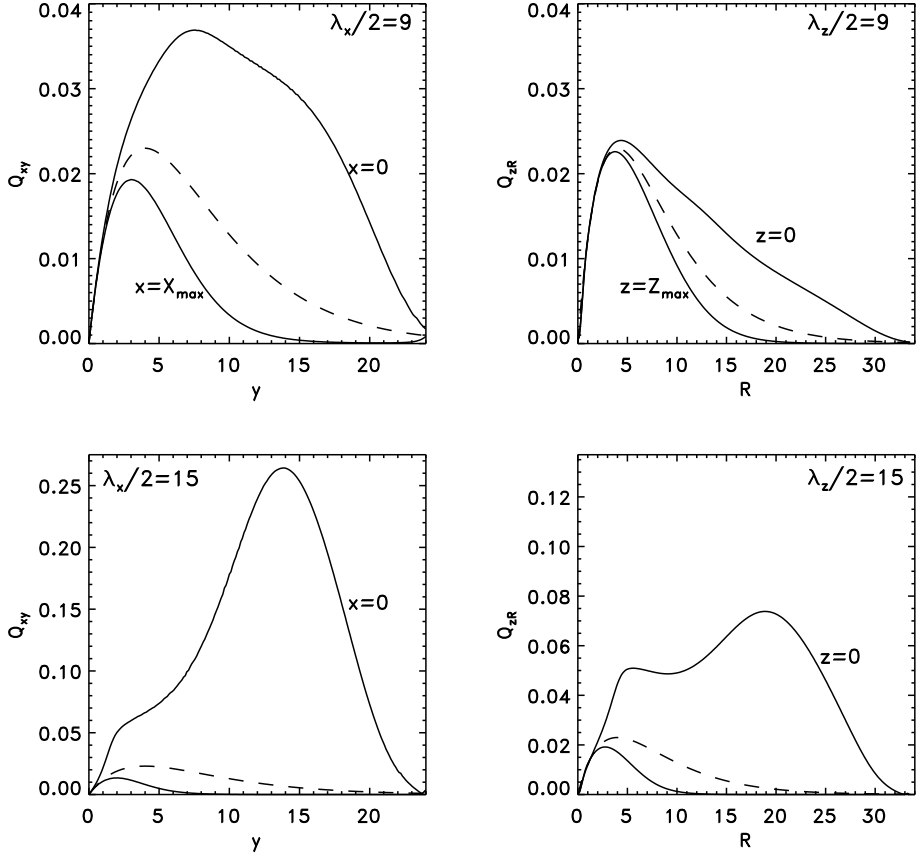}
  \caption{The dependence of the functions 
$Q_{xy}$ and $Q_{zR}$ on $y$ and $R$, respectively in the initial
states (dashed lines) and in the final states (solid lines), with
$\alpha=1$ and constant gravity. 
In the left panels, $Q_{xy}$ is shown along the lines $x=0$ and
$x=X_{\rm max}$ for two different wavelengths: $\lambda_{x}/2=9$ and $15$.
The right panels display $Q_{zR}$ along the lines $z=0$ and $z=Z_{\rm max}$
for $\lambda_{z}/2=9$ and $15$, as quoted at the corners of these panels.
  }
  \label{fig:QQ}
\end{figure*}

The source term of the differential equation in the initial
cylindrical configuration with constant $g_{R}$ is:
\begin{equation}
Q_{zR}^{(i)}= \frac{1}{8(1+\alpha)^{3}}R\exp\left(-\frac{R}{2(1+\alpha)}\right).
\label{eq:rhs_cyl}
\end{equation}
By comparing Eqs.~(\ref{eq:rhs_2D}) and (\ref{eq:rhs_cyl}), 
we see that $Q_{zR}^{(i)}$ 
has a similar dependence on $R$ as $\hat{Q}_{xy}^{(i)}$
on $y$, except by the factor $R/(2+2\alpha)$. As a consequence,
$Q_{zR}^{(i)}$ has a maximum at $2(1+\alpha)=2H$ whereas
$\hat{Q}_{xy}^{(i)}$ is a monotonically decreasing function of $y$.
In order to facilitate comparison between both cases, it is convenient
to use $Q_{xy}$ defined as $Q_{xy}\equiv y\hat{Q}_{xy}/(2+2\alpha)$ 
because both $Q_{xy}$ and $Q_{zR}$ are identical at the initial state.
However, they are expected to depart one from each 
other in the nonlinear equilibrium state because of the changes
introduced by the different geometries.
The form of $Q_{xy}$ and $Q_{zR}$ in the final equilibrium states
will be discussed in the next Section.

\section{Final equilibrium states}
\label{sec:final}
The final equilibrium state can be reached from the initial configuration
through continuous deformations of field lines using a iterative method
as described in M74. To do so,
we add a perturbation to the magnetic variables
in the initial equilibrium state,
\begin{equation}
\delta A(x,y)=-\mu A_{0}(y) \sin \left(\frac{\pi y}{Y_{\rm max}}\right)
\cos \left(\frac{\pi x}{X_{\rm max}}\right),
\end{equation}
for the Cartesian model and
\begin{equation}
\delta \Phi (z,R)=-\mu \Phi_{0}(R) \sin \left(\frac{\pi R}{R_{\rm max}}\right)
\cos \left(\frac{\pi z}{Z_{\rm max}}\right),
\end{equation}
for the axisymmetrical model. Here $\mu$ is the amplitude of the
perturbation which we chose between $0.02$ and $0.06$.
Note that the perturbations are symmetric about $y=0$ in the 
Cartesian model\footnote{The Cartesian model
allows also solutions where field lines cross the midplane. They
are called odd-parity solutions or midplane antisymmetric modes.}.

Our main interest is to compare the final equilibrium states in
Cartesian and axisymmetrical models using the same external gravitational
potential along $y$ and $R$, respectively. In this way, the comparison 
can be done on a common ground. In the next subsection, we will do so
for models with constant gravitational acceleration.
A more realistic gravitational potential
will be studied in Section \ref{sec:nonuniform} in order to 
examine how it might affect the final nonlinear states.

\subsection{Constant gravitational acceleration}
\label{sec:constantacc}
Parker (1966) showed that a plane-parallel atmosphere with uniform gravity
and a horizontal magnetic field is unstable to the undular
mode if the horizontal wavelength of the perturbation
is larger than a critical value, namely,  $\lambda_{x}>\lambda_{\rm crit}
\equiv 4\pi (1+\alpha) (2\alpha+1)^{-1/2}$.  For the reference
value $\alpha=1$,
the dimensionless critical wavelength is $14.5$ or, equivalently, $7.26H$.
Magnetic tension can stabilize disturbances with shorter wavelengths. 
The horizontal wavelength of the mode with the maximum growth rate 
depends on the vertical wavelength. For a typical value of 
$Y_{\rm max}=24$, the maximum
growth rate occurs at $X_{\rm max}=12$ (i.e.~$\lambda_{x}=24$).

The stability of magnetized cylinders has been considered in the
past (e.g., Stodolkiewicz 1963; Nagasawa 1987) but, to our knowledge,
none derived the dispersion relation under the assumption
of constant gravitational field. Nevertheless, 
the linearized equations of motion under axial geometry and
axisymmetric perturbations are identical to eqs.~(III-3)-(III-6)
quoted in Parker (1966), once the subscripts $y$ and $z$ in
Parker's notation are replaced for $z$ and $R$,  except for the 
curvature term $v_{R}/R$ that appears as an extra term
in the continuity and energy equations.
Since the curvature term becomes small at scales comparable or
larger than $\lambda_{\rm crit}$,
the critical wavelength is essentially the same in both geometries.
In fact, we have checked that our iterative scheme converges to the 
initial state as long as $\lambda_{z}\lesssim 7.5H$ (when $\alpha=1$
is used), indicating that
they are stable.

In the Cartesian atmosphere, M74 found that if the upper boundary is far enough
from the galactic plane, i.e.~$Y_{\rm max}\gg (1+\alpha)$ in
dimensionless units, the evolution
of most of the matter in the domain is not affected by the location
of the upper boundary.
The reason is that more than $90\%$  of the energy of the initial
state resides under $y<3.5(1+\alpha)$ (see M74).
In the axisymmetrical model, we can apply the same reasoning 
to expect that the solution is not significantly altered as long
as the box size is much larger than the effective scaleheight,
defined as the radius $R$ at which $\Phi_{0}$
decays by a factor $e$, namely, $4.3(1+\alpha)$ [note that the effective
scaleheight of  $A_{0}$ in the Cartesian model is $2(1+\alpha)$]. 
In Figure \ref{fig:hbox} we compare the the magnetic flux of the
final state in the axisymmetrical model 
using $\alpha=1$ and $\lambda_{z}/2=12$,
when the upper boundary, $R_{\rm max}$, is located at $24$ and at $33.5$. 
We see that both maps are almost identical at $R<20$.
The main difference is that upper magnetic field lines in the magnetic
bulge region
inflate further out when the boundary is far away because a larger
column of material is unloaded and brought down to the valley of the curved
magnetic field lines.
Therefore, this upper region becomes lighter and more buoyant.
By contrast, cuts of $\Phi$ along the valley 
$z=0$ are very similar in both cases (see the lower panel 
of Fig.~\ref{fig:hbox}). This means that the extra material that
has been loaded into the valley when the upper boundary is placed
at $33.5$, does not cause significant additional compression
in the radial direction.

\begin{figure}
  \includegraphics[width=1.7\columnwidth]{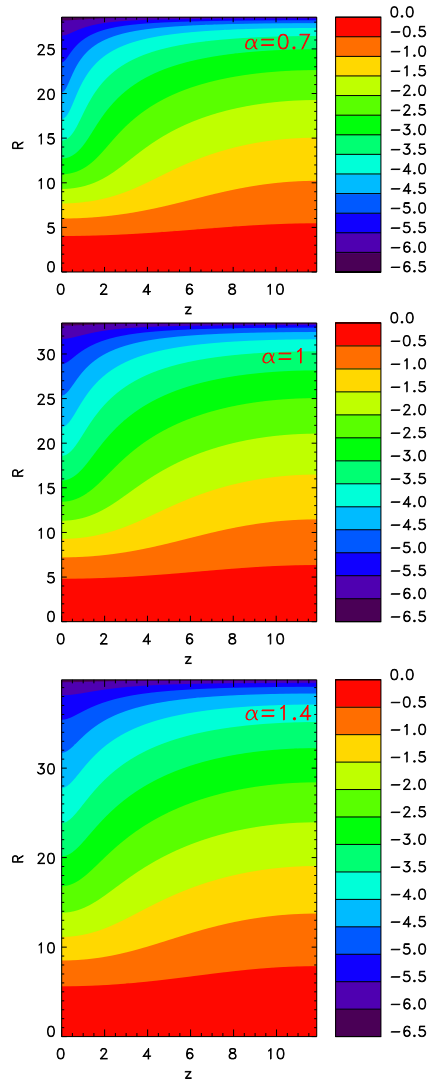}
\vskip 0.4cm
  \caption{Color map of the magnetic flux $\Phi(z,R)$ in natural logarithmic
scale for the final state for 
$\lambda_{z}/2=12$ (in dimensionless units) and different values of $\alpha$.  
The radial gravitational acceleration $g_{R}$ was assumed to be constant.  }
  \label{fig:alpha}
\end{figure}

\begin{figure*}
  \includegraphics[width=2.0\columnwidth]{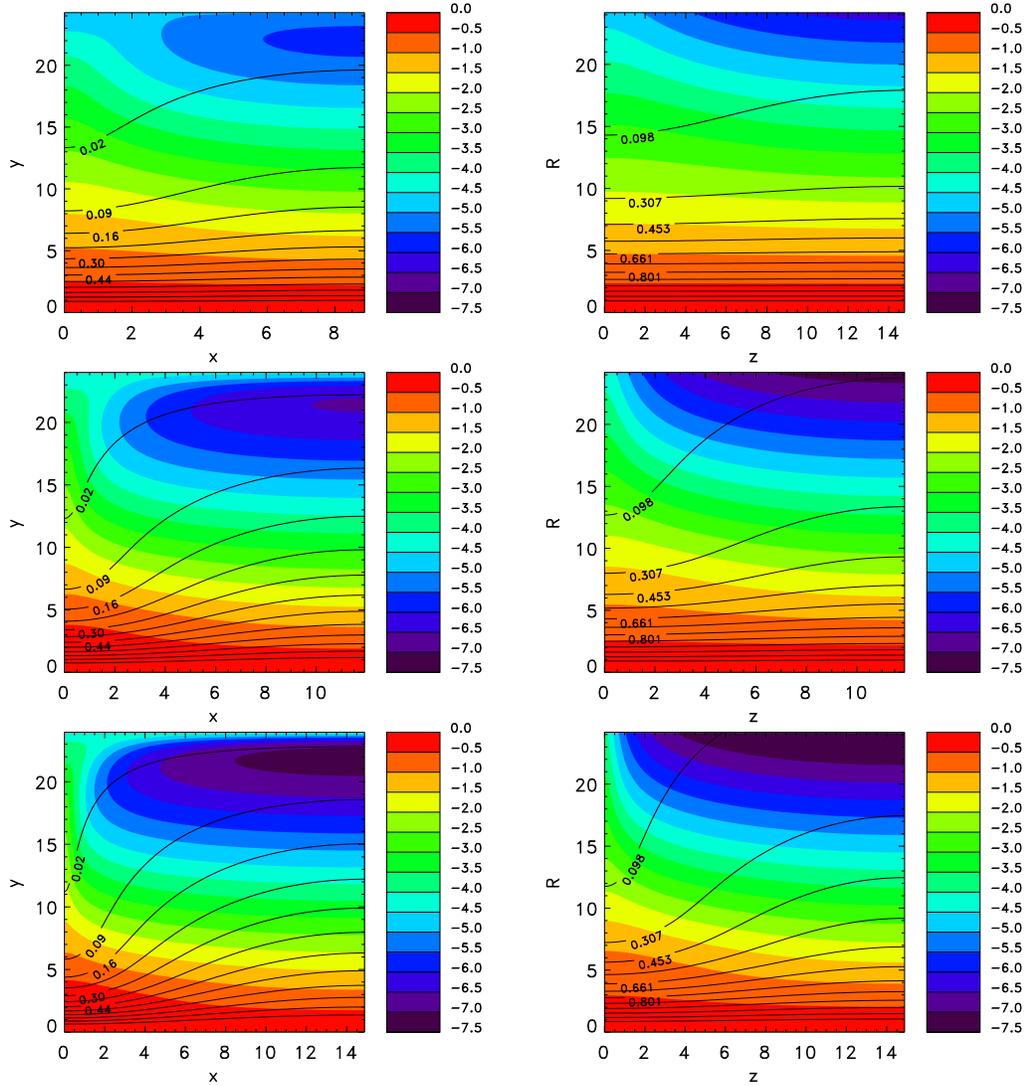}
\vskip 0.4cm
  \caption{Contour plots of the density (color) and magnetic field lines
(solid lines) for the Cartesian model (left) and the axisymmetric model (right),
under constant gravity,
for half the wavelength of $9,12$ and $15$ in units of $c_{s}^{2}/g$ (from
top to bottom) and $\alpha=1$. 
The color bar gives the correlation between the colors
and the natural logarithm of the density.
The number on some curves is the value of $A$ in the Cartesian model or the
value of $\Phi$ in the axisymmetric model. To make the comparison
of the deformation of the magnetic field lines easier, 
we plot the same magnetic lines in both models.
  }
  \label{fig:2D_cyl}
\end{figure*}

In order to have the
same dynamical range in all our calculations, we will place our
upper boundary at the distance at which the variable of interest
($A$ for the Cartesian model and $\Phi$ for the axisymmetric case) decreases
to $\exp (-6.12)=0.0022$ of its maximum value. 
With this convention, $Y_{\rm max}=12.2(1+\alpha)$ and
$R_{\rm max}=16.7(1+\alpha)$. Interestingly, in the axisymmetric case,
the corresponding density at $R_{\rm max}$ 
is only $5.4\times 10^{-8}$ of its value at $R=0$.

Different magnetostatic equilibrium configurations can be
found by choosing the function $q(A)$ for the Cartesian
model and $q(\Phi)$ for the axisymmetric one. Parker (1966)
explored the cases in which $q(A)$ is either a linear or
quadratic function of $A$. In practice,
these functions are not free because they are defined by
the initial conditions.
M74 emphasized that $q$ increases when $A$ decreases in
the Cartesian model. In Fig.~\ref{fig:qA}, the functions $q(A)$ for
the Cartesian layer and $q(\Phi)$ for the axially-symmetric configuration
are plotted for $\alpha=1$ and $\lambda_{x}/2=\lambda_{z}/2=15$.
In the initial state, the function $q_{0}(A)$
follows a power-law with index $-2\alpha$ (see Section \ref{sec:Cartesian}
and Fig.~\ref{fig:qA}).  
In the final equilibrium state,
the power-law index varies with $A$, but $q(A)$ is
always a monotonically decreasing function. As discussed in
Section \ref{sec:axisymmetric} and shown in Fig.~\ref{fig:qA}, the slope
of $q$ in the
initial state of the axisymmetric model is steeper than
in the Cartesian model. In fact, $q_{0}$ reaches a maximum value of $10^{7.3}$
in the axisymmetric model, but only $10^{6.2}$ in the Cartesian model.
Note that the range of the $q$-axis in Fig.~\ref{fig:qA} was taken somewhat
different in order to facilitate comparison of the shape of the curves.
The behaviour of $q$ versus $\Phi$ in the final equilibrium state
is similar as $q$ versus $A$ (see Fig.~\ref{fig:qA}), although
the new geometry introduces some remarkable differences 
in the shape of the function
$q(\Phi)$. For instance, the change of $q$, relative
to its initial value, in the range $0.6<\log_{10}\Phi^{-1}<1.5$
is smaller than in the same range of $A$ in the Cartesian model.
The similar appearance of the functions $q(A)$ and $q(\Phi)$
should not lead us to the false conclusion that the source terms
of the elliptic differential equations, namely $Q_{xy}(x,y)$
and $Q_{zR}(z,R)$, are also similar. Figure \ref{fig:QQ}
displays $Q_{xy}$ at two different cuts along $x=0$ and $x=X_{\rm max}$,
and $Q_{zR}$ along $z=0$ and $z=Z_{\rm max}$.
The fractional change of $Q_{xy}$ from its initial
value is always larger than the change of $Q_{zR}$.
Both $Q_{xy}$ and $Q_{zR}$ may present two local maxima
along a cut through the maximum
heights of the magnetic field lines, i.e.~along 
$x=X_{\rm max}$ and $z=Z_{\rm max}$.
The relative amplitude of these maxima varies with the
wavelength of the perturbation. 

Figure \ref{fig:alpha} shows contour plots of the magnetic flux in
the final axisymmetric state for a disturbance
with $\lambda_{z}/2=12$ and three different $\alpha$ values
($\alpha=0.7,1.0$ and $1.4$).
We see that the maps are a scaled version to each other by the factor
$1+\alpha$. Since it is common for astrophysical purposes to assume
equipartition between magnetic field and gas pressures, we will restrict
ourselves to the case $\alpha=1$ hereafter. 

\begin{figure}
  \includegraphics[width=1.12\columnwidth]{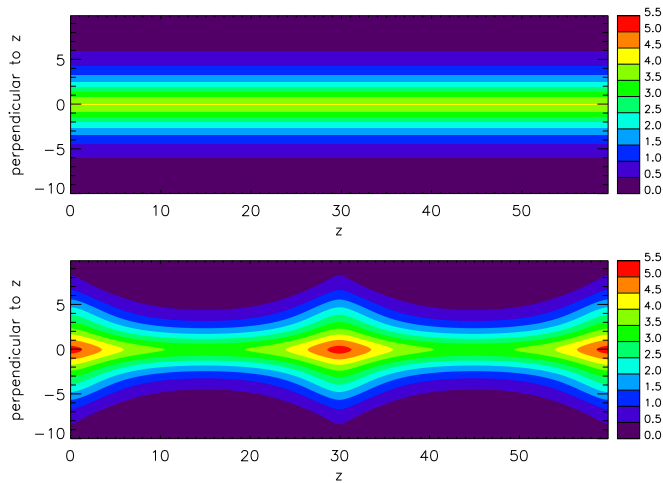}
\vskip 0.4cm
  \caption{Column density map for the initial (top panel)
and final axisymmetric state (bottom panel) along a line of sight
perpendicular to the axis of symmetry ($z$-axis). We used
$\alpha=1$, $\lambda_{z}/2=15$ and uniform radial 
gravity. The units of surface density
and length are $\rho_{0}(0)c_{s}^{2}/g$ and 
$c_{s}^{2}/g$, respectively. Note that the scale is linear.

  }
  \label{fig:column}
\end{figure}

Figure \ref{fig:2D_cyl} exhibits contour plots of the density (color)
and magnetic field lines in the final stable state of the axisymmetric model
for three different horizontal wavelengths ($\lambda_{z}/2=9, 12$ and $24$).
For comparison, density and magnetic field lines are also shown for
the Cartesian model.
Although the vertical size of the computational domain is larger in the
axisymmetric case, we show the same box in all the cases for ease 
in visualization. When the systems are disturbed with the same wavelength,
it is clear from Fig.~\ref{fig:2D_cyl} that the field lines and the
density contours become more
deformed in the Cartesian model than under axial symmetry.
It turns out that a cylindrical initial configuration is more rigid
to perturbations and the magnetic field is less buoyant
than a plane-parallel layer.
Interestingly, the same degree of deformation of magnetic
field lines found for $\lambda_{x}/2=9$ in the Cartesian model, can be achieved
in the axisymmetric case but for a longer wavelength of $\lambda_{z}/2=12$.
Likewise, the level of radial buoyancy of the magnetic field for 
$\lambda_{z}/2=15$ is comparable to the level
of vertical buoyancy for $\lambda_{x}/2=12$ in the Cartesian model.

\begin{figure}
  \includegraphics[width=1.0\columnwidth]{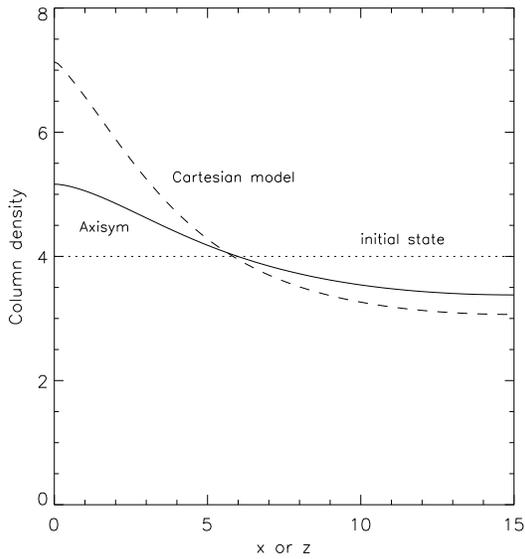}
\vskip 0.4cm
\caption{Column density for the final states as a function of the
horizontal coordinate for $\lambda/2=15$ and constant gravity. 
In the Cartesion model, the column density
is $\Sigma(x)=\int_{-\infty}^{\infty} \rho (x,y)dy$. In the
axisymmetric model, we plot the column density along a line of
sight perpendicular to the axis of symmetry and with null impact
parameter, i.e.~$\Sigma(z)=\int_{-\infty}^{\infty}\rho(z,R) dR$.
The column density at the initial states is drawn as a dotted line.}
\label{fig:den_axis}
\end{figure}

The column density map of a filament with $\alpha=1$ and
whose axis of symmetry lies in the
plane of the sky is given in Fig.~\ref{fig:column} in the initial and
final equilibrium states with $\lambda_{z}/2=15$. 
The scaleheight of the column density along a condensation in
the direction perperdicular to $z$ is about twice the
scaleheight in the wings of the condensations.
As a consequence of the less efficient drainage of gas into the magnetic
valleys, the column density enhancement in the gas 
condensations is rather modest. In fact,
the maximum enhancement in the column density 
over a line of sight with zero impact parameter,
is smaller than in the Cartesian model (see Fig.~\ref{fig:den_axis}).
The column density may increase up to a factor of $7/4$ in Cartesian
geometry (see also Kim et al.~2004) but only by a factor of $5/4$ in the
axisymmetric model.

The result that a cylindrical configuration is more rigid
than a plane-parallel layer has, {\it a posteriori}, a simple geometrical 
explanation 
as follows. Since magnetic forces act only perpendicular to the
field lines, gas pressure gradients must balance the gravitational
forces along a field line regardless the geometry of the problem. 
The plasma must slide down along the magnetic field lines until it reaches
a new configuration of pressure equilibrium.
The point is that, because of the radial convergence of the flow 
streamlines in the axisymmetrical model, we need less displacement 
of gas in the 
axisymmetric case than in the Cartesian model to achieve the same
pressure gradient, as soon as the gas is isothermal, and, 
henceforth, less deformation of the magnetic field lines.

\begin{figure*}
  \includegraphics[width=2.0\columnwidth]{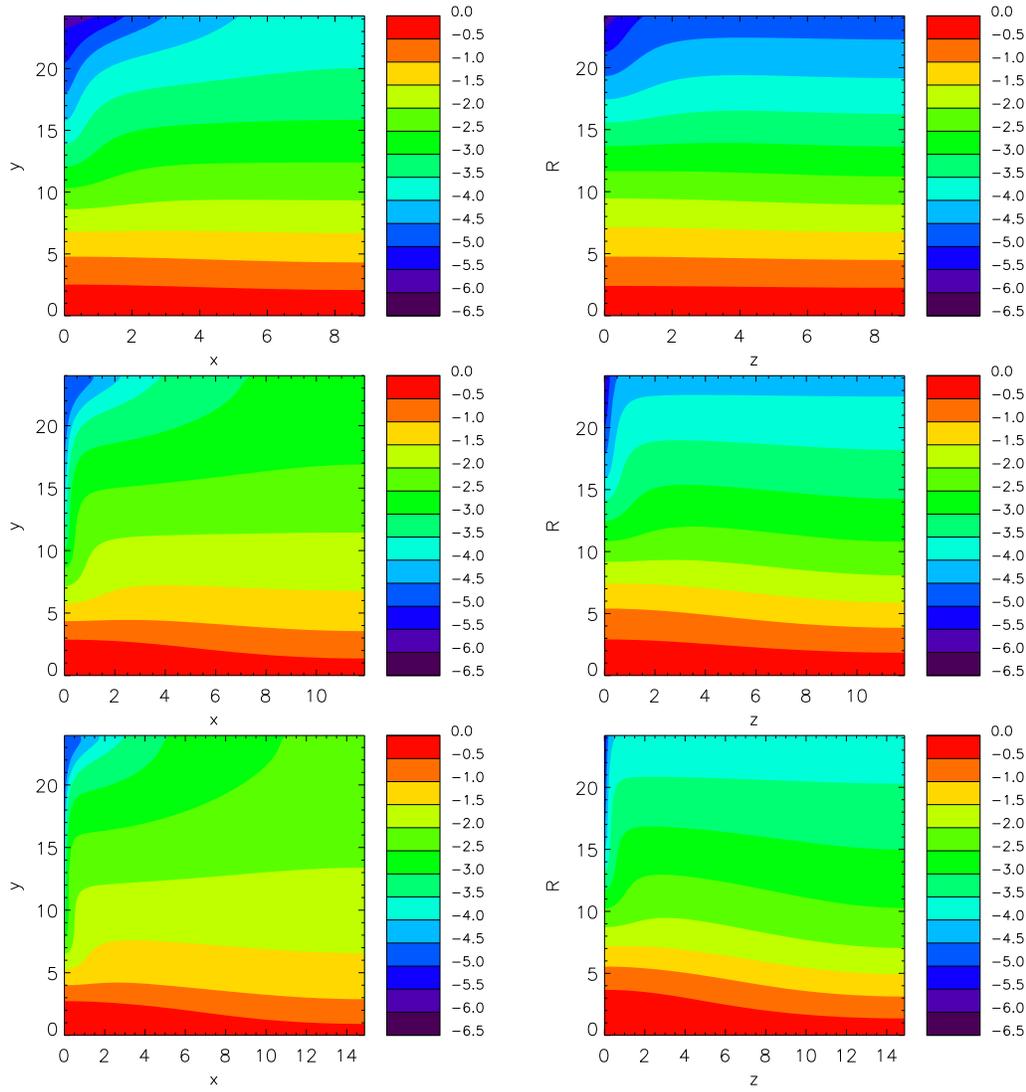}
\vskip 0.4cm
  \caption{Magnetic pressure maps, normalized to its value on the $z$-axis
in the initial state, in the six final states of Figure \ref{fig:2D_cyl}.
  }
  \label{fig:mag}
\end{figure*}

Since the separation between magnetic field lines in azimuthal
sections do not reflect the magnetic strength
in the axisymmetric case, it is illustrative to compare the magnetic pressure
to make a fair comparison between the results in Cartesian and axial
geometries (see Fig.~\ref{fig:mag}).
Given a certain wavelength of the perturbation,
the magnetic pressure maps depend on the geometry.
Because of the stronger deformation in the Cartesian
scenario, the magnetic pressure force and the magnetic
stresses along the valleys of the magnetic field lines, i.e.~at the
cut $x=0$, are larger than along $z=0$. 
The distribution of magnetic
pressure in the axial model with $\lambda_{z}/2=12$ 
is quite similar to the magnetic pressure configuration in the 
Cartesian case with $\lambda_{x}/2=9$.

\begin{figure}
  \includegraphics[width=0.64\columnwidth]{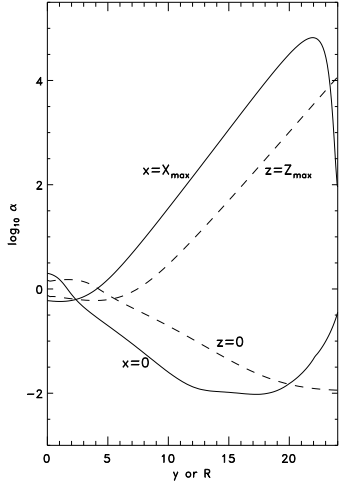}
\vskip 0.4cm
  \caption{
Decadic logarithm of the ratio between the magnetic and
gas pressures in the final state under constant gravity condition. 
The two solid lines give $\alpha$
values along the lines $x=0$ and $x=X_{\rm max}=15$ for the Cartesian model 
and the dashed lines give $\alpha$ along $z=0$ and $z=Z_{\rm max}=15$
in the axisymmetric case. 
  }
  \label{fig:alphafinal}
\end{figure}

The ratio of the magnetic to gas pressures is an indicator of the
efficiency with which gas flows along field lines.
As the increase of gas density
in the magnetic field valleys is due primarily to the drainage along
the field lines (see also M74), 
we expect that if the drainage is less efficient in the axisymmetrical
case, the variation of
$\alpha$ in the final state relative to the initial values should be
also smaller. In Fig.~\ref{fig:alphafinal}, we plot the distribution of the 
magnetic-to-gas pressure ratio along $x=0$ and $x=X_{\rm max}$ for
the Cartesian model, and along the same cuts $z=0$ and $z=Z_{\rm max}$ 
for the axisymmetric model
in the final state, for a perturbation with
half the wavelength of $15$. Although the general shape of the curves is
rather similar in both geometries, the $\alpha$ values in the range
$10<y<22$ at $x=X_{\rm max}$ are almost a factor $10$ larger
than in the axisymmetric model at the same region 
(i.e.~$10<R<22$ and $z=Z_{\rm max}$).

\subsection{Non-uniform gravitational acceleration}
\label{sec:nonuniform}
In a infinite self-gravitating isothermal filament, 
the gravitational acceleration $g_{R}$ at large $R$ decays as $1/R$, 
whereas it increases linearly at small $R$.
Therefore, the assumption that $g_{R}$ is constant is only
valid at some intermediate range in $R$.
To close our analysis, in this subsection we present the final 
equilibria for axially-symmetric configurations in a more realistic 
external gravitational field. 

According to the results of Kim \& Hong (1998) for a Cartesian
model, the growth time is reduced by almost an order of magnitude 
and the length scale of the most unstable mode by factors of $4$--$8$ 
when a more realistic gravity model is used because
in such a realistic disk the magnetic field strength decreases
rapidly with vertical distance which promotes the development
of strong magnetic buoyancy.
We expect, therefore, that the length and timescales of the Parker
instability in a filament will be modified if a more realistic 
gravity model is adopted.

In order to have a model as self-consistent as possible, we will
assume that the external gravitational potential is dominated by a collisionless
(stellar or dark matter) component with isothermal one-dimensional
velocity dispersion $\sigma_{\star}$. The gravitational potential
created by an isothermal filament is given by 
\begin{equation}
\psi(R)=2\sigma_{\star}^{2} \ln \left(1+\frac{R^{2}}{8H_{\star}^{2}}\right),
\end{equation}
where $H_{\star}$ is a parameter that specifies the scale height of
the collisionless component satisfying $4\pi G \rho_{0,\star}(0)H_{\star}^{2}=
\sigma_{\star}^{2}$, where $\rho_{0,\star}(0)$ is the density
of collisionless matter at $R=0$ (e.g., Ostriker 1964). 
As already anticipated, the gravitational acceleration
$g_{R}(R)$ increases linearly with $R$ at $R\ll \sqrt{8}H_{\star}$,
reaches a maximum at $\sqrt{8}H_{\star}$ 
and decreases as $1/R$ at $R\gg \sqrt{8}H_{\star}$.
Under this gravitational field, the density and magnetic
field in the unperturbed state are:
\begin{equation}
\frac{\rho_{0}(R)}{\rho_{0}(0)}=\frac{B_{0}^{2}(R)}{B_{0}^{2}(0)}=
\left(1+\frac{R^2} {8H_{\star}^{2}}\right)^{-s},
\end{equation}
where $s$ is defined by $s\equiv 2\sigma_{\star}^{2}/[(1+\alpha)c_{s}^{2}]$.
If we define the scale height of the gas, $H$, as the radial distance 
at which the density of the gas is reduced to 
$1/e$ of its value at $R=0$, then $H^{2}=8(\exp(1/s)-1)H_{\ast}^{2}$.

\begin{figure}
  \includegraphics[viewport=30 0 324 279]{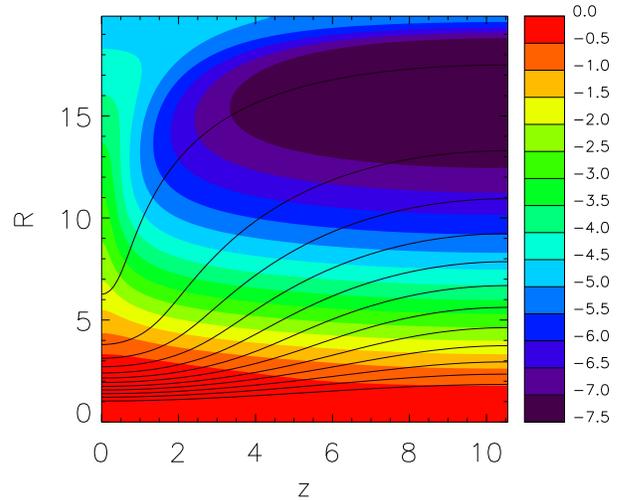}
\vskip 0.4cm
\caption{Distribution of the volume density and
magnetic field lines in the axisymmetric final equilibrium state 
in a non-uniform radial
gravitational field with $s=4$. 
In order to facilitate comparison with Fig.~\ref{fig:2D_cyl} in which $H=2$,
distance is measured in units of $H/2$. 
Field lines (solid curves) were chosen so that 
they have the same magnetic strength
in the initial state than those displayed in Fig.~\ref{fig:2D_cyl}.}
\label{fig:realistic1}
\end{figure}

\begin{figure}
  \includegraphics[width=1.12\columnwidth]{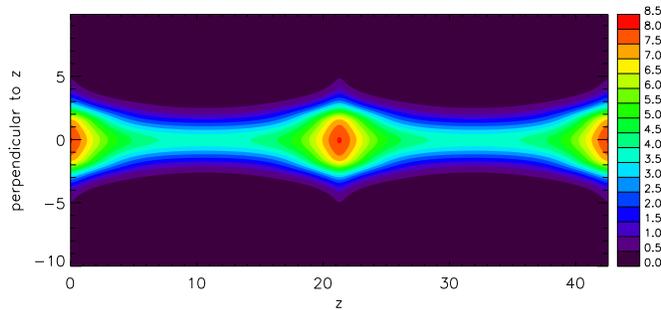}
\vskip 0.4cm
  \caption{Column density map for the 
final axisymmetric state along a line of sight
perpendicular to the axis of symmetry ($z$-axis). We used
$\alpha=1$, $s=4$ and $\lambda_{z}=21.3$. The units of density
and length are $\rho_{0}(0)$ and 
$H/2$, respectively. Note that the scale is linear.

  }
  \label{fig:column_ost}
\end{figure}

To illustrate the effect of using a non-constant gravitational acceleration,
we useed $\alpha=1$ but different values of $s$.
We ran models where the upper cap was located at the distance $R_{\rm max}$ 
at which the density decreases to $(1-5)\times 10^{-6}$ of its value at $R=0$.
We have empirically derived that the larger the $s$-value is, 
smaller the critical wavelength becomes. 
For $s=4$, perturbations with $\lambda_{z}/2=3.85H_{\star}=2.55H$ 
are marginally unstable, whereas they become marginally
unstable at $\lambda_{z}/2=1.5H$ when $s=8$.
In terms of $H$, this result implies that the critical wavelength 
for $s=4$--$8$ is smaller
by almost a factor of $1.5$--$2.5$ than that for the model with constant 
gravity (see Section \ref{sec:constantacc}) and, thereby, the
growth timescale is shorter for larger $s$-values.
Figure \ref{fig:realistic1} shows the final equilibrium state 
for $s=4$ and $\lambda_{z}/2=5.3H$. This $\lambda_{z}$-value is close 
to the horizontal wavelength with maximum growth rate. 
By comparing with Figure \ref{fig:2D_cyl}, we
note that although the global appearance is similar to the case with
constant gravity and $\lambda_{z}/2=15$, 
the field lines appear slightly more deformed. 
The main effect of a non-constant gravitational acceleration strength is
to shorten the separation of the magnetic valleys.
The projected surface density that an observer would see if the axis of
the filament lies in the plane of the sky is shown in Fig.~\ref{fig:column_ost} 
for a model with $s=4$.
The enhancement in the column density
over a line of sight passing through the core of
a condensation is $2.1$ in this model. This factor is appreciably
larger than that obtained in axisymmetric models with constant
gravity but similar to that found in Cartesian models with
uniform gravity.

\section{Discussion and conclusions}
\label{sec:conclusions}
The Parker instability dictates that longitudinal magnetic
field lines that give some support against gravity
are always unstable to undular perturbations. It has long
been realized that the Parker
instability, which is interesting in itself, may play
a significant role to understand the formation of massive
clouds in the interstellar
medium, probably aided by thermal and gravitational instabilities
(Mouschovias et al.~2009). 
The Parker instability can also play a role in the evolution
of filamentary structure, such as filamentary clouds, at least at the
early stages of the instability. 
While the properties of the final states are well documented for Cartesian
models (e.g., M74; Basu et al.~1997; Kim et al.~2001), it was unclear
how they depend on the adopted geometry.
In order to fill this gap, our primary goal was to investigate the nonlinear
outcome of such instability in an axisymmetric model where the
initial equilibrium configuration consists of a infinite cylinder
in the presence of a longitudinal magnetic field. 
We quantified the level of buoyancy of the magnetic field and 
drainage of the gas to promote density enhancements,
as compared to the Cartesian models,
which are currently being used to study
the formation of substructure in the interstellar medium.
We focused on understanding the physics
of the Parker instability in filaments, 
rather than on making detailed comparisons with observations. 
This may be very useful as a first step to interpret full numerical 
simulations.

In the Cartesian model, the potential vector function $A$ obeys
a nonlinear elliptic equation, while that for the axisymmetric model
is the magnetic flux $\Phi$. However, the differential operator
as well as the source terms are different. 
In order to gain some insight on the different nature of the equations,
we have compared the form of $q(A)$ and $q(\Phi)$ at the initial
and final states in each model.
The source terms depend directly on the derivatives of these functions.
However, in spite that the aspect of these functions looks like similar
at a glance, the source terms $Q_{xy}$ and $Q_{zR}$
turn out to be absolutely different in the final state. This is
a consequence of the nonlinearity of the differential equations.

We have focused first on models with uniform external gravity.
We found that the axial model is more rigid than the Cartesian layer,
in the sense that the magnetic field is less buoyant
and the drainage of gas less efficient. In fact, 
in the axisymmetric model the matter shrinks in radius and
generates a convergent flow that facilitates to 
achieve pressure equilibrium with less deformation of field lines.
In order to have the same level of deformation of the field lines
than in the Cartesian model with half the wavelength of $9c_{s}^{2}/g$,
we need a perturbation of half the wavelength of $12 c_{s}^{2}/g$,
for $\alpha=1$. Under uniform gravity, we find that a factor $5/4$ enhancement 
in column density can be obtained in the axisymmetric model for
the more unstable wavelength, which is modest as compared to
the factor of $7/4$ found in the Cartesian model.
At the position of the wings of the condensations, the ratio
of the magnetic-to-gas pressures
becomes larger than $1$ beyond a certain radius because
of the evacuation of gas in these regions.
However, it is still approximately $10$ times smaller than in
Cartesian geometry.

We have examined a more realistic choice to describe the variation
of the external gravitational potential in axisymmetric models. 
The main effects of
a non-uniform gravitational acceleration is to modify the separation of
magnetic valleys and thereof the timescale of the Parker instability.
A factor of $2$ enhancement in column density over its initial equilibrium
value can be found in axisymmetric models with non-uniform radial gravity. 
Hence, unless this density enhancement is sufficient to initiate thermal
or gravitational instabilities, the Parker instability should be thought 
of as setting the stage upon which other small-scale processes will be 
modeling the structure and evolution of the filament.

\acknowledgements{We would like to thank Jos\'e Franco and Jongsoo Kim for
helpful comments and the anonymous referee for very valuable
suggestions, which have led to substantial improvements
to the manuscript.
The authors acknowledge financial support from CONACyT
project CB-2006-60526 and PAPIIT project IN121609.}

\begin{thebibliography}
\bibitem[Alfaro et al.(1992)]{alf92} Alfaro, E. J., Cabrera-Ca\~no, J., 
\& Delgado, A. J. 1992, \apj, 399, 576
\bibitem[Ass\'eo et al.(1978)]{ass78} Ass\'eo, E., Cesarsky, C. J., 
Lachi\`eze-Rey, M., \& Pellat, R.  1978, \apj, 225, L21
\bibitem[Ass\'eo et al.(1980)]{ass80} Ass\'eo, E., Cesarsky, C. J., 
Lachi\`eze-Rey, M., \& Pellat, R.  1978, \apj, 237, 752
\bibitem[Basu et al.(1997)]{bas97}
Basu, S., Mouschovias, T. Ch., \& Paleologou, E. V. 1997, \apj, 480, L55
\bibitem[Blitz \& Shu(1980)]{bli80}
Blitz, L., \& Shu, F. H. 1980, \apj, 238, 148
\bibitem[Conselice et al.(2001)]{con01}
Conselice, C. J., Gallagher III, J. S., \& Wyse, R. F. G. 2001, 
\apj, 122, 2281
\bibitem[Dungey(1953)]{dun53}
Dungey, F. W. 1953, \mnras, 113, 180
\bibitem[Franco et al.(2002)]{fra02}
Franco, J., Kim, J., Alfaro, E. J., \& Hong, S. S. 2002, \apj, 570, 647
\bibitem[Genzel \& Stutzki(1989)]{gen89}
Genzel, R., \& Stutzki, J. 1989, \araa, 27, 41
\bibitem[Giz \& Shu(1993)]{giz93}
Giz, A. T., \& Shu, F. H. 1993, \apj, 404,185
\bibitem[Hanasz \& Lesch(2000)]{han00}
Hanasz, M., \& Lesch, H. 2000, \apj, 543, 235
\bibitem[Hanawa et al.(1992)]{han92}
Hanawa, T., Matsumoto, R., \& Shibata, K. 1992, \apj, 393, L71
\bibitem[Jackson et al.(2010)]{jac10}
Jackson, J. M., Finn, S. C., Chambers, E. T., Rathborne, J. M.,
\& Simon, R. 2010, \apj, 719, L185
\bibitem[Kim et al.(2000)]{kim00}
Kim, J., Franco, J., Hong, S. S., Santill\'an, A., \& Matos, M. A. 2000,
\apj, 531, 873
\bibitem[Kim \& Hong(1998)]{kim98}
Kim, J., \& Hong, S. S. 1998, \apj, 507, 254
\bibitem[Kim et al.(1997)]{kim97}
Kim, J., Hong, S. S., \& Ryu, D. 1997, \apj, 485, 228
\bibitem[Kim et al.(2004)]{kim04}
Kim, J., Ryu, D., \& Hong, S. S. 2004, \apss, 292, 255
\bibitem[Kim et al.(2001)]{kim01}
Kim, J., Ryu, D., \& Jones, T. W. 2001, \apj, 557, 464 
\bibitem[Kosi\'nski \& Hanasz(2006)]{kos06}
Kosi\'nski, R., \& Hanasz, M. 2006, \mnras, 368, 759
\bibitem[Kosi\'nski \& Hanasz(2007)]{kos07}
Kosi\'nski, R., \& Hanasz, M. 2007, \mnras, 376, 861
\bibitem[Lachi\`eze-Rey et al.(1980)]{lac80}
Lachi\'eze-Rey, M., Cesarsky, C. J., Ass\'eo, E., \& Pellat, R. 1980,
\apj, 238, 175
\bibitem[Lee \& Hong(2007)]{lee07}
Lee, S. M., \& Hong, S. S. 2007, \apjs, 169, 269
\bibitem[Mouschovias(1974)]{mou74}
Mouschovias, T. Ch. 1974, \apj, 192, 37
\bibitem[Mouschovias(1976)]{mou76}
Mouschovias, T. Ch. 1976, \apj, 206, 753
\bibitem[Mouschovias(1996)]{mou96}
Mouschovias, T. Ch. 1996, in Solar and Astrophysical Magnetohydrodynamic
Flows, vol.~281, ed.~K.~C.~Tsinganos, NATO ASI Ser.~C (Kluwer, Dordrecht),
p.~9 
\bibitem[Mouschovias et al.(2009)]{mou09}
Mouschovias, T. Ch., Kunz, M. W., \& Christie, D. A. 2009, \mnras, 397, 14
\bibitem[Mouschovias et a.(1974)]{mou74b}
Mouschovias, T. Ch., Shu, F. H., \& Woodward, P. R. 1974, \aap, 33, 73 
\bibitem[Nagasawa(1987)]{nag87}
Nagasawa, M. 1987, Progress of Theoretical Physics, 77, 635
\bibitem[Nakamura et al.(1991)]{nak81}
Nakamura, F., Hanawa, T., \& Nakano, T. 1991, \pasj, 43, 685
\bibitem[Nakamura et al.(1995)]{nak95}
Nakamura, F., Hanawa, T., \& Nakano, T. 1995, \apj, 444, 770
\bibitem[Nakano(1979)]{nak79}
Nakano, T. 1979, \pasj, 31, 697
\bibitem[Ostriker(1964)]{ost64}
Ostriker, J. P. 1964, \apj, 140, 1056
\bibitem[Parker(1966)]{par66}
Parker, E. N. 1966, \apj, 145, 811
\bibitem[Ryu et al.(1998)]{ryu98}
Ryu, D., Kang, H., \& Biermann, P. L. 1998, \aap, 335, 19
\bibitem[Salom\'e et al.(2006)]{sal06}
Salom\'e, P. et al. 2006, \aap, 454, 437
\bibitem[Santill\'an et al.(2000)]{san00}
Santill\'an, A., Kim, J., Franco, J., Martos, M., Hong, S. S., \& Ryu, D.
2000, \apj, 545, 353
\bibitem[Shu(1974)]{shu74}
Shu, F. H. 1974, \aap, 33, 55
\bibitem[Stodolkiewicz(1963)]{sto63}
Stodolkiewicz, J. S. 1963, Acta Astronomica, 13, 30
\bibitem[Tomisaka et al.(1988)]{tom88}
Tomisaka, K., Ikeuchi, S., \& Nakamura, T. 1988, \apj, 335, 239
\end{thebibliography}
\end{document}